# Towards the surface hydroxyl species in CeO$_2$ nanoparticles

Tatiana V. Plakhova[a], Anna Yu. Romanchuk[a], Sergei M. Butorin[b], Anastasia D. Konyukhova[a], Alexander V. Egorov[a], Andrey A. Shiryaev[a,c], Alexander E. Baranchikov[a,d], Pavel V. Dorovatovskii[e], Thomas Huthwelker[f], Evgeny Gerber[g,h], Stephen Bauters[g,h], Madina M. Sozarukova[d], Andreas C. Scheinost[g,h], Vladimir K. Ivanov[a,d], Stepan N. Kalmykov[a,c] and Kristina O. Kvashnina[g,h] *

a. *Lomonosov Moscow State University, Department of Chemistry, Leninskije Gory 1, Moscow, Russia.*
b. *Molecular and Condensed Matter Physics, Department of Physics and Astronomy, Uppsala University, P.O. Box 516, Uppsala, Sweden*
c. *Frumkin Institute of Physical Chemistry and Electrochemistry of Russian Academy of Science, Moscow, Russia*
d. *Kurnakov Institute of General and Inorganic Chemistry, Russian Academy of Sciences, Moscow, Russia*
e. *National Research Centre "Kurchatov Institute", Moscow, Russia*
f. *Paul Scherrer Institute, Swiss Light Source, WLGA 211, CH-5232, Villigen, The Switzerland*
g. *The Rossendorf Beamline at ESRF – The European Synchrotron, CS40220, 38043 Grenoble Cedex 9, France*
h. *Helmholtz Zentrum Dresden-Rossendorf (HZDR), Institute of Resource Ecology, PO Box 510119, 01314 Dresden, Germany.*

\* Corresponding author: Kristina O. Kvashnina kristina.kvashnina@esrf.fr

Understanding the complex chemistry of functional nanomaterials is of fundamental importance. Controlled synthesis and characterization at the atomic level is essential to gain deeper insight into the unique chemical reactivity exhibited by many nanomaterials. Cerium oxide nanoparticles have many industrial and commercial applications, resulting from very strong catalytic, pro- and anti-oxidant activity. However, the identity of the active species and the chemical mechanisms imparted by nanoceria remain elusive, impeding the further development of new applications. Here, we explore the behavior of cerium oxide nanoparticles of different sizes at different temperatures and trace the electronic structure changes by state-of-the-art soft and hard X-ray experiments combined with computational methods. We confirm the absence of the Ce(III) oxidation state at the surface of CeO$_2$ nanoparticles, even for particles as small as 2 nm. Synchrotron X-ray absorption experiments at Ce L$_3$ and M$_5$ edges, combined with X-ray diffraction (XRD), high-resolution transmission electron microscopy (HRTEM) and small angle X-ray scattering (SAXS) and theoretical calculations demonstrate that in addition to the nanoceria charge stability, the formation of hydroxyl groups at the surface profoundly affects the chemical performance of these nanomaterials.

## Introduction

Nanotechnology is a rapidly growing field, with one of the most active areas being the development of nanomedical products to develop new therapeutic strategies against cancer. Cerium oxide possesses excellent pro- and anti-oxidant properties because of the action of the Ce(III)/Ce(IV) redox couple at its surface.[1-5] Approaching the nanoscale, ceria catalysts become profoundly more reactive due to the enhanced surface area to volume ratio, indicating that the interface chemistry is crucial. Many studies have indicated that the shape and size of nanoparticles (NPs) considerably affects the selectivity and activity of reactions involving CeO$_2$ NPs. The ability to understand the NP surface science is vital, as it allows interactions between NPs and surrounding environments to be characterized and understood, and finally also to be tuned to obtain even higher efficacy.

It has been repeatedly shown that CeO$_2$ NPs actively participate in processes occurring in biological media at ambient temperatures,[6,7] with the performance of the NPs dictated by NP size. CeO$_2$ NPs with sizes larger than 10 nm cause oxidative stress, which leads to the formation of reactive oxygen species (ROS) such as hydrogen peroxide, superoxide radical, hydroxyl radical and so forth in living cells.[8-10] Conversely, ceria NPs of 2-10 nm typically act as anti-oxidants by ROS scavenging. However, the size effect is not solely responsible for ceria efficiencies and several other factors have to be taken into account. These factors include particle shape, surface chemistry, and surface modifiers or ligands that can participate in complexation and redox reactions.[11-14] It is commonly believed that the behavior of ceria NPs is closely related to the changes in oxygen non-stoichiometry, e.g., the Ce(IV)/Ce(III) ratio.[8, 15-18]

The dependence of complex reactivity on the ceria oxidation state and the presence of Ce(III) species at the surface have been studied earlier, using multiple analytical techniques, including X-ray photoelectron spectroscopy (XPS),[10, 19-22] resonant photoemission spectroscopy (RPES),[23, 24] X-ray absorption spectroscopy (XAS),[25-28] electron energy loss spectroscopy (EELS),[29, 30] electron magnetic resonance spectroscopy (EMS)[31] and chemical methods.[32] Unfortunately, high-vacuum conditions and X-ray exposure can affect the Ce(IV)/Ce(III) ratio in CeO$_2$ NPs. Zhang and co-authors[27] demonstrated that the concentration of Ce(III) determined by XPS increases dramatically after specimen are stored overnight in a vacuum chamber (from 11.6% immediately after sample-loading to 29.3%). Our previous work[33] also confirmed the partial reduction of Ce(IV) to Ce(III) during XPS measurements. In turn, in the majority of papers in which XAS has been used for ceria NP characterization,[25-27] the Ce(III) concentration does not exceed 5% (i.e., close to the lower detection limit) and is independent of particle size. These results suggest that the differences in biological activity of ceria NPs of various origins cannot be explained simply by variations in Ce(III) content.

Despite the large body of experimental evidence showing the beneficial effects of nanoceria in living systems, the underlying key chemical mechanisms and the nature of the active species responsible for these outstanding properties remains elusive. This mainly results from the likelihood of relatively low concentrations of the active species and the inability to distinguish the contributions from surface and sub-surface species and structures. This requires direct, advanced characterization and theoretical validation. Here, we use X-ray diffraction (XRD), high-resolution transmission electron microscopy (HRTEM), small angle X-ray scattering (SAXS), high energy resolution fluorescence detected hard X-ray absorption spectroscopy (HERFD-XAS) and soft X-ray XAS techniques to elucidate the structure, surface and chemical states of monodispersed ceria NPs varying in diameter from 2 to 8 nm. Soft X-ray spectroscopy at the Ce $M_5$ edge probes 4f states,[34] while HERFD-XAS at the Ce $L_3$ edge interrogates the 5d orbitals, which are involved in chemical bonding[35, 36] (Fig. S1). Additionally, Ce $L_3$ edge measurements yield complementary information concerning the number of 4f electrons through quadrupole-allowed electronic transitions.[25, 37] Coupling these two techniques, we show that spectral profiles at the Ce $L_3$ and Ce $M_5$ edges of ceria NPs exhibit different features. The results are compared to those of bulk $CeO_2$ (particles larger than 25 nm). These experiments confirm the absence of Ce(III) at the surface of all NPs, while revealing the influence of the surface on the electronic structure of ceria NPs. Experimental results are supported by state-of-the-art ab-initio calculations and theory based on the single Anderson impurity method (SIAM). Furthermore, the HERFD-XAS studies at the Ce $L_3$ edge demonstrate how conventional drying procedures affect the surface of nanoceria, which in turn modifies the physical and chemical properties of the resulting nanomaterials.

Figure 1 illustrates the processes involved in the HERFD-XAS and soft XAS experiments that are key to the results. The energy diagrams and electronic transitions, which involve the creation of core holes in the 3d and 2p states, thereby probing the Ce 4f and Ce 5d valence shells, are also shown in Figure 1.

## Experimental section

### Synthesis route

Aqueous solution of cerium (III) nitrate hexahydrate was used as stock solutions. To prepare the solutions, a calculated amount of cerium salt was dissolved in MilliQ water (18.4 MOhm/cm). Final concentrations of $Ce(NO_3)_3 \cdot 6H_2O$ were determined to be 0.001M, 0.1M and 0.8M. For NP precipitation, the cerium stock solution was added to a fivefold excess of 3M aqueous ammonia at room temperature under continuous stirring. The precipitate formed was allowed to age for 12 h. After that the yellow precipitate was separated by centrifugation and washed three times with MilliQ water to remove soluble admixtures. For further characterization, $CeO_2$ samples were prepared in three different ways: "as-prepared sample" – $CeO_2$ NPs after washing procedure without any thermal treatment; "sample dried at 40 °C" – as-prepared NP after drying overnight under ambient environment at 40 °C; "sample dried at 150 °C" – as-prepared NP after drying overnight under an ambient environment at 150 °C.

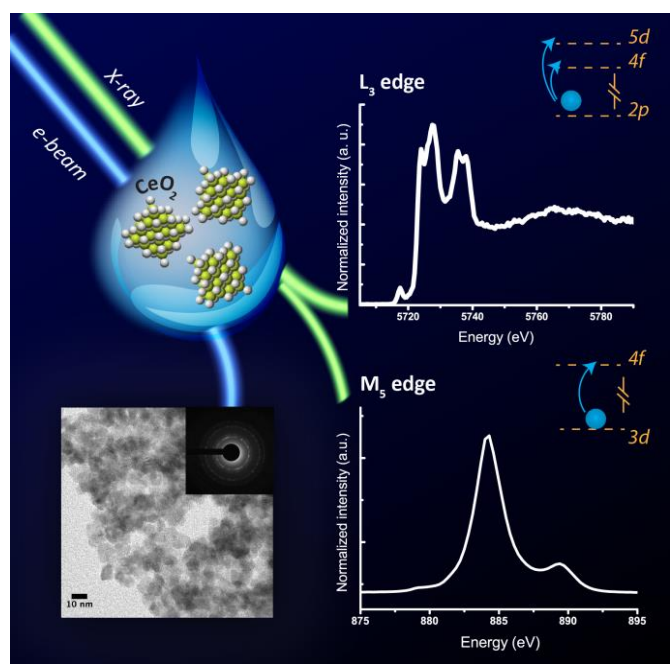

Figure 1. Illustration of the advanced experimental probes and processes. Valence and surface features of ceria NPs were studied by high resolution X-ray absorption techniques: HERFD-XAS for $L_3$ edge absorption and soft XAS for $M_5$ edge. Together with the spectroscopic investigation of NPs, they were examined with HRTEM (bottom left) for a comprehensive analysis.

### Characterizations of $CeO_2$ NPs

The XRD data were recorded using a Bruker D8 Advance diffractometer (Cu Kα radiation). The PC- PDF database was used for the identification of the crystalline phases. No traces of amorphous phases in the XRD pattern were detected. The synchrotron-based XRD measurements were performed at the "XSA (X-ray Structural Analysis)" beamline of the Kurchatov Synchrotron Radiation Source (NRC "Kurchatov Institute", Moscow). For synchrotron-based XRD measurements $CeO_2$, NPs samples were placed in synthetic vacuum oil and mounted on 20 μm nylon CryoLoop. The measurements were performed in the transmission mode using a Rayonix SX-165 CCD detector at a wavelength of 0.80 Å. Raw 2D scattering images were integrated using the Fit2D program.

Microstructural evaluation studies were performed using JEOL-2100F transmission electron microscopes (HRTEMs) with accelerating voltages of 200 kV. Electron diffraction patterns were recorded using the same instruments.

Small-angle X-ray scattering patterns were registered using monochromatic Cu Kα radiation in a broad angular range (scattering vectors(s) between 0.1 and 27 nm$^{-1}$), using a SAXSess diffractometer (Anton Paar). Samples were placed in standard X-ray capillaries. A Kratki collimation scheme was used. Slit geometry providing a highly intense primary X-ray beam was employed; reduction to point geometry (desmearing) was performed using a standard algorithm.[38] All measurements were performed at room temperature in an evacuated chamber (residual pressure 5–10 mbar).

HERFD-XAS experiments were performed at beamline BM20 of the European Synchrotron Radiation Facility in Grenoble. The incident energy was selected using the <111> reflection from a double Si crystal monochromator. Rejection of higher harmonics was achieved by two Si mirrors working at an angle of 2.5 mrad relative to the incident beam. The incident X-ray flux was ~10$^{10}$ photons·s$^{-1}$ at the sample position. HERFD-XAS spectra were measured using an X-ray emission spectrometer[39] at 90° horizontal scattering angle. Sample, analyzer crystal and photon detector (avalanche photodiode) were arranged in a vertical Rowland geometry. The Ce HERFD-XAS spectra at the L$_3$ edge were obtained by recording the maximum intensity of the Ce Lα$_1$ emission line (4839 eV) as a function of the incident energy. The emission energy was selected using the <331> reflection of one spherically bent Ge crystal analyzer (with R = 1m) aligned at a 87° Bragg angle. The size of the beam at the sample was 400 μm horizontal times 150 μm vertical. A combined (incident convoluted with emitted) energy resolution of 0.9 eV was obtained, as determined by measuring the FWHM of the elastic peak. Samples for the HERFD-XAS measurements were prepared as wet pastes and sealed with single kapton confinement.

The ceria M$_5$ XAS edge spectra were measured at the PHOENIX beamline located at the Swiss Light Source (SLS) in Villigen, Switzerland. The beamline covers soft and tender X-rays ranging from 0.4-8 keV, by use of two branchlines: A low energy branchline serves 0.4-2 keV and a high energy branchline serves 0.8-8 keV. The source of the beamline is an elliptical undulator. Both branchlines provide either an unfocussed beam or a micro-focus of about 3 μm. The measurements presented here were made on the high energy branchline. This branch employs a double crystal monochromator and a high harmonics suppressor consisting of 2 planar mirrors. To generate monochromatic light at energies about 0.8 keV, a Beryl crystal was used. The incoming flux ($I_0$) was measured from the total electron yield taken from a polyester foil (0.5 μm thickness), which was coated with 50 nm of nickel. The beamsize used in the experiments was 700 μm vertical times 900 μm horizontal. Both total fluorescence and total electron yield were measured on the sample. As fluorescence yields are low at M$_{4,5}$ edges, only the total electron yield data were used. A Keithley picoammeter was used to collect the current from a $I_0$ signal and sample.

### Theoretical simulations

The Ce L$_3$ theoretical calculations were performed using the FEFF 9.6. code.[40] The input files were based on the atomic unit cell parameter a=0.541 nm. Calculations were made in a similar way, as reported by Li et al.[41] previously for bulk CeO$_2$. Full multiple scattering (FMS) calculations were performed using a Hedin-Lundqvist self-energy correction and other standard cards. The apparent reduction of the core hole lifetime broadening was achieved by using the EXCHANGE card. The spectrum for the CeO$_2$ 2 nm NPs with OH groups has been calculated for a cluster of 315 atoms. The structure was optimized using an Avogadro program.

The calculations of the Ce M$_5$ edge spectra were performed in a framework of the Anderson impurity model.[42] The spectra of bulk CeO$_2$ were calculated in a manner described in Ref.[34] for the Ce(IV) system, taking into account the Ce 4f hybridization with the valence states and the full multiplet structure, due to intra-atomic interactions. The ground (final) state of the spectroscopic process was described by a linear combination of the 4f$^0$, 4f$^1\underline{v}^1$ and 4f$^2\underline{v}^2$ (3d$^9$4f$^1$, 3d$^9$4f$^2\underline{v}^1$ and 3d$^9$4f$^3\underline{v}^2$) configurations, where $\underline{v}$ stands for an electronic hole in the valence level. The values for the model parameters were set the same as in Ref.[34]: the energy for the electron transfer from the valence band to the unoccupied Ce 4f level Δ=2.0 eV; the 4f-4f Coulomb interaction U$_{ff}$=9.0 eV; the 3d core hole potential acting on the 4f electron U$_{fc}$=12.6 eV and the Ce 4f – valence state hybridization term V=1.0 eV (0.6 eV) in the ground (final) state of the spectroscopic process.

## Results and discussion

The present study solidly identifies, for the first time, the active species of ceria NPs by applying experimental X-ray spectroscopy methods and theoretical calculations. Nanoceria of different sizes were prepared using a facile, rapid chemical precipitation method[33]. The CeO$_2$ NPs obtained were characterized by XRD, HRTEM and SAXS methods to determine size, morphology, and chemical composition. Based on the analytical results (Fig. S2, S3, S4), NP diameters were 2 nm for CeO$_2$ prepared from 0.001M Ce(NO$_3$)$_3$, 5 nm for CeO$_2$ prepared from 0,1M Ce(NO$_3$)$_3$ and 8 nm for CeO$_2$ prepared from 0.8M Ce(NO$_3$)$_3$.

Figure 2 shows the HERFD-XAS experimental spectra (red curves) of CeO$_2$ NPs with 2 nm, 5 nm and 8 nm diameters recorded at the Ce L$_3$ edge, in comparison to the spectrum of bulk CeO$_2$ (particles > 25 nm in diameter, green curve). Green curves are identical in all panels of Figure 2 and are plotted here for clarity of results. Measurements were carried out at the Ce 2p core level, which means that the electron being excited by the incident X-rays is from the 2p orbital of Ce at the energy of 5723 eV. It was excited into an empty valence orbital of CeO$_2$ NPs.

The spectra reveal the presence of the dipole-allowed 2p-5d transitions (main edge transitions in the energy range 5720-5730 eV) and of the quadrupole 2p-4f transitions (at the pre-edge in the energy range 5715-5720 eV). These experiments utilized an X-ray emission spectrometer[39] to record transitions with high-energy resolution that reveal the fine structure of the 5d and 4f orbitals. This allows us to discern and compare the spectral features and structures originating at the surface (in this case for 2 nm NPs) and from the bulk of the ceria.

In accordance with previous investigations,[25] the pre-edge structure for the three sizes of NPs reveals solely the presence of the Ce(IV) oxidation state (Fig. S5). This is in agreement with the previously observed absence of Ce(III) inside $CeO_2$ NPs, even with sizes around 2 nm[25] without any post-synthesis treatment. In contrast, the 5d electronic states in HERFD-XAS spectra (or main edge transitions) have different shapes.[26, 43] We found that peaks in the $L_3$ HERFD-XAS spectra of 2 nm $CeO_2$ are essentially just starting to develop and are broad in comparison to those obtained from 5 nm, 8 nm and bulk $CeO_2$. This would be in line with electron delocalization at the surface of NPs, and hence there is a decreasing probability of discrete electron transfer from the 2p to the 5d orbital. Note that it is the cerium 5d orbitals which are chemically active and can form additional bonds with precursor or surfactant materials. Therefore, we investigate here the impact of hydroxyl groups on the electronic structure of nanoceria

.

The nature of hydroxylation of the $CeO_2$ surface is still an open question. Previously, the formation of OH groups onto the $CeO_2$ surface as a result of water dissociation was observed for samples prepared by vapor chemical deposition.[44, 45] In the present work, the $CeO_2$ was synthesized from water solution; therefore the presence of cerium hydroxide phase is also possible. Badri et al.[46] investigated $CeO_2$ obtained by a similar synthesized method, using infrared spectroscopy, and observed the presence of OH groups that disappeared during calcination. The authors explain the observed OH groups as evidence of "cerium oxyhydroxide microphases". The XRD data in the current research (Fig.S7) confirm that nanoparticle samples (as prepared) consist of a pure $CeO_2$ phase; the presence of amorphous hydroxide layers cannot be fully excluded, however, by XRD measurements. The OH groups on the metal oxide surface will form in water or under water vapor by protonation of the surface oxygen groups, depending on the pH. In the present study, the aqueous suspension of the as-prepared $CeO_2$ samples has a near neutral pH, i.e., slightly below the point of zero charge for $CeO_2$ (8.1).[47] Therefore, formation of surface OH groups is, in fact, likely for the non-dried samples.

With the help of ab-initio calculations (the blue curves in Figure 2), it is shown that spectral shape changes because of the presence of OH-groups at the surface of NPs, pointing towards the strong influence of the surface composition on the reactivity and performance of nanomaterials in specific applications. Theoretical calculations for bulk $CeO_2$ have been performed using an identical procedure, as reported by Li et al.[41] Similar to the work of Li et al., we show here only part of the absorption spectra, which corresponds to the 2p-5d transitions, and omit multi-electron excitations from the experimental data, which appear at higher incident energy (reported in Figure 1 in the incident energy range 5736-5738 eV). The theoretical spectrum for 2 nm $CeO_2$ (Figure 2 top, lower blue curve) has been calculated using the same input file and parameters as for bulk $CeO_2$, but with a cluster size corresponding to 2 nm particles. A noticeable difference between the two calculations is evident. More drastic changes occur, however, once the oxygen surface groups of the 2 nm NPs are hydroxylated at the surface of the 2 nm $CeO_2$ NPs (Fig. 2 top, upper blue curve denoted 2 nm theory with OH). As soon as particles grow large, the effects of the OH groups at the surface are diminished in the spectra, but are still visible in the case of 5 nm $CeO_2$ and disappear completely for particles as large as 8 nm. The theoretical results show that changes in the HERFD-XAS spectral profiles originate from the different distribution of the unoccupied Ce 5d states near the Fermi levels in three different cases (bulk $CeO_2$, 2nm $CeO_2$ and 2nm $CeO_2$ with OH groups).

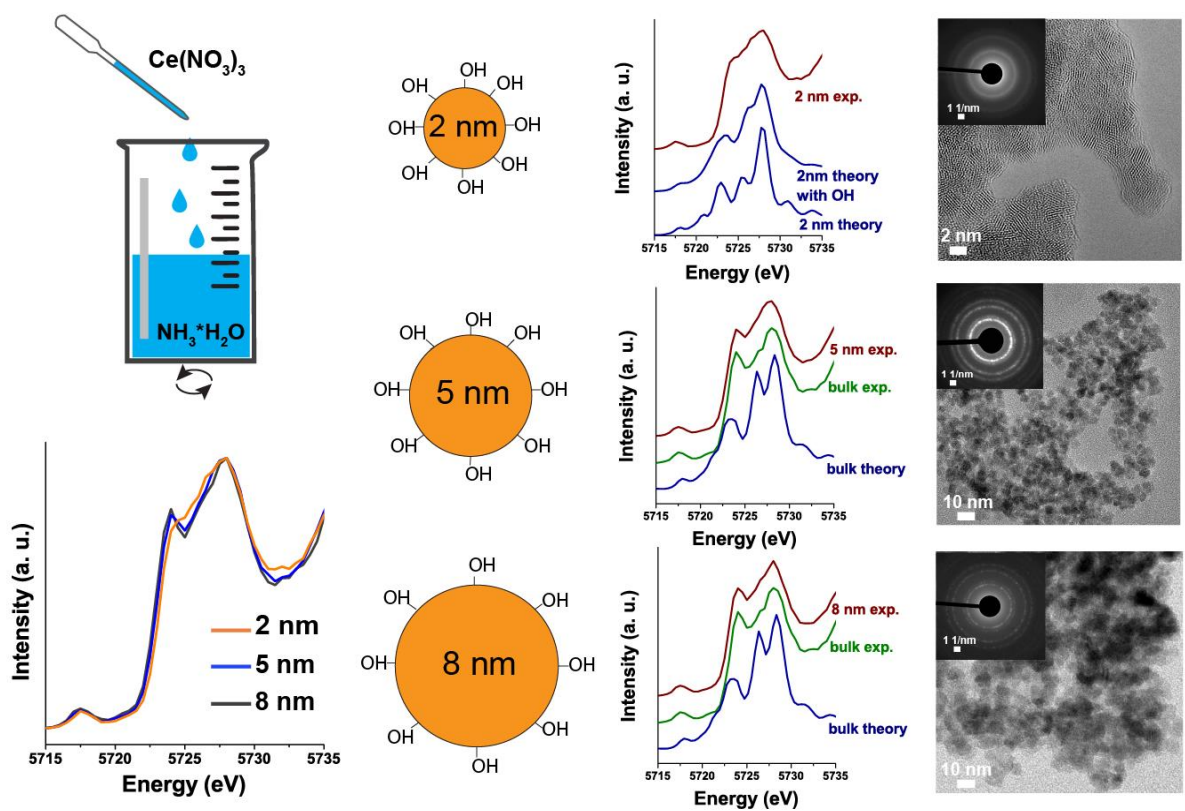

Figure 2. Schematic drawing of the synthesis of $CeO_2$ NPs and conceptual depictions of NPs structure with hydroxyl groups at the surface. Representative Ce HERFD-XAS spectra for 2 nm; 5 nm; 8 nm $CeO_2$ NPs recorded at the Ce $L_3$ edge compared to the bulk $CeO_2$ (> 25 nm) and theoretical calculations. The HRTEM images of $CeO_2$ NPs of 2 nm, 5 nm and 8 nm size and selected area electron diffraction (SAED) data (upper left inserts) are also shown.

Besides providing new insight into the structure of NPs at the surface, these results pose new questions to the field of chemistry and, on a more technical level, on the coupling mechanism of X-ray excited states to the surface electronic structure and the localization of the valence electrons in very small NPs.

**Counting electrons on NPs surfaces**

Structural insights on the surface layers present in 2 nm and 5 nm NPs were obtained by soft XAS experiments. The results provided by the HERFD-XAS experiments were confirmed by measurements in the soft X-ray energy range (~800 eV) at the Ce $M_5$ edge, as reported in Figure 3. An excited 4f state, which is coupled to the ground 3d state, reveals the absence of Ce(III) at the surface of NPs. The spectra of the $CeO_2$ NPs at the Ce $M_{4,5}$ edges were previously described[48] as consisting of surface and bulk components.[49] The contribution of the surface component increases with decreasing NP sizes.

Our measurements at the Ce $L_3$ edge do not confirm the presence of any Ce(III) in our samples. Figure 3 shows that the shape of the Ce $M_5$ spectra is very similar for all $CeO_2$ NPs and can be well reproduced by model calculations for the Ce(IV) system, which take into account the Ce 4f - O 2p hybridization and the full multiplet structure, due to intra-atomic interactions at the Ce site. The spectra exhibit a distinct and strong high-energy satellite at ~5.5 eV above the main line.

This high-energy satellite has been discussed before for bulk $CeO_2$ (e.g. Ref. [34]) and shown to be due to Ce 4f - O 2p charge-transfer, as a result of strong Ce 4f - O 2p hybridization and configurational mixing. A good agreement of the calculations with the experiments suggests that contributions of the $4f^0$, $4f^1\underline{v}^1$ and $4f^2\underline{v}^2$ configurations in the ground state amount to 49%, 46% and 5%, respectively. The Ce 4f occupancy in the ground state was found to be 0.56 electrons, which indeed indicates a highly covalent character of the chemical bonding, as in bulk $CeO_2$. The soft XAS approach reconciles previous disparities in the results obtained by different spectroscopic methods and importantly acknowledges that NPs show identical electronic structure information by soft and hard X-ray spectroscopy if they are performed at identical conditions.

**Post-processing temperature treatment of NPs**

We have previously discussed the measurement results and theoretical calculations for the samples obtained immediately after synthesis (as-prepared samples). To further elucidate the effect of surface hydroxylation on the electronic structure of the NPs, we subjected them to a 24-h heat treatment at 40 °C and 150 °C in air. According to XRD and HRTEM measurements (Fig. 4, Fig. S7, S8), the drying process did not affect the phase crystallinity, morphology, and particle size distribution. In contrast, the spectral shapes at the Ce $L_3$ edge changed when going from as-prepared to dried $CeO_2$ NPs, suggesting changes of the electronic structure.

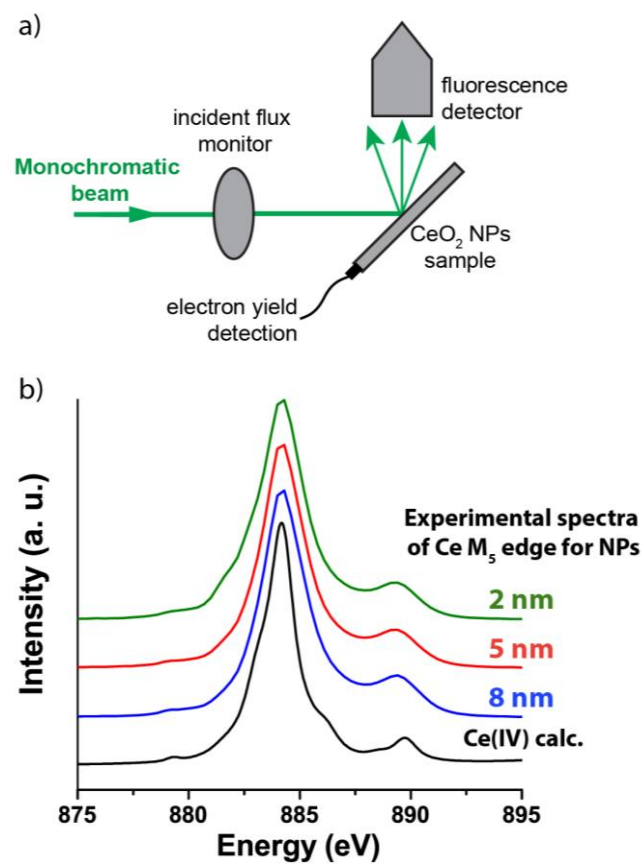

Figure 3. Schematic of the experimental setup for XAS measurements at the Ce $M_5$ edge (a). Spectra of 2, 5 and 8 nm $CeO_2$ NPs (green, red and blue curves) have been measured in total-electron-yield mode and compared to the calculations for $CeO_2$ using the SIAM (black curves).

Figure 4 shows that drying 2 nm $CeO_2$ at 40 °C and 150 °C resulted in a change of the main Ce $L_3$ edge transitions corresponding to the excitations to the Ce 5d orbitals. The intensity difference of the 2p-5d transitions between three ceria samples (as-prepared; dried at 40 °C and 150 °C) is pronounced in the case of small NPs. For the 5 nm NPs, drying still induces weak spectral changes, while no effect of temperature treatment is observed for the 8 nm NPs. As expected, changes of the electronic structure in 2 and 5 nm $CeO_2$ NPs are associated with a dehydroxylation of the NP surface: Upon thermal treatment, ceria NPs lose OH groups from their surface and the effect is more pronounced for small NPs. According to thermogravimetric analysis coupled with a mass spectrometry for 2 nm NPs, sample water is released in the temperature range 40 – 200°C (Fig.S8). Surface contributions from the NPs structure decrease with the increase leads to significant changes and a decrease in surface hydroxylation. This leads us to assume that the antioxidant activity of the particles is directly related to concentration of hydroxyl species at the surface: with an increase of surface area to volume ratio, the antioxidant activity of NPs increase. Our in size.

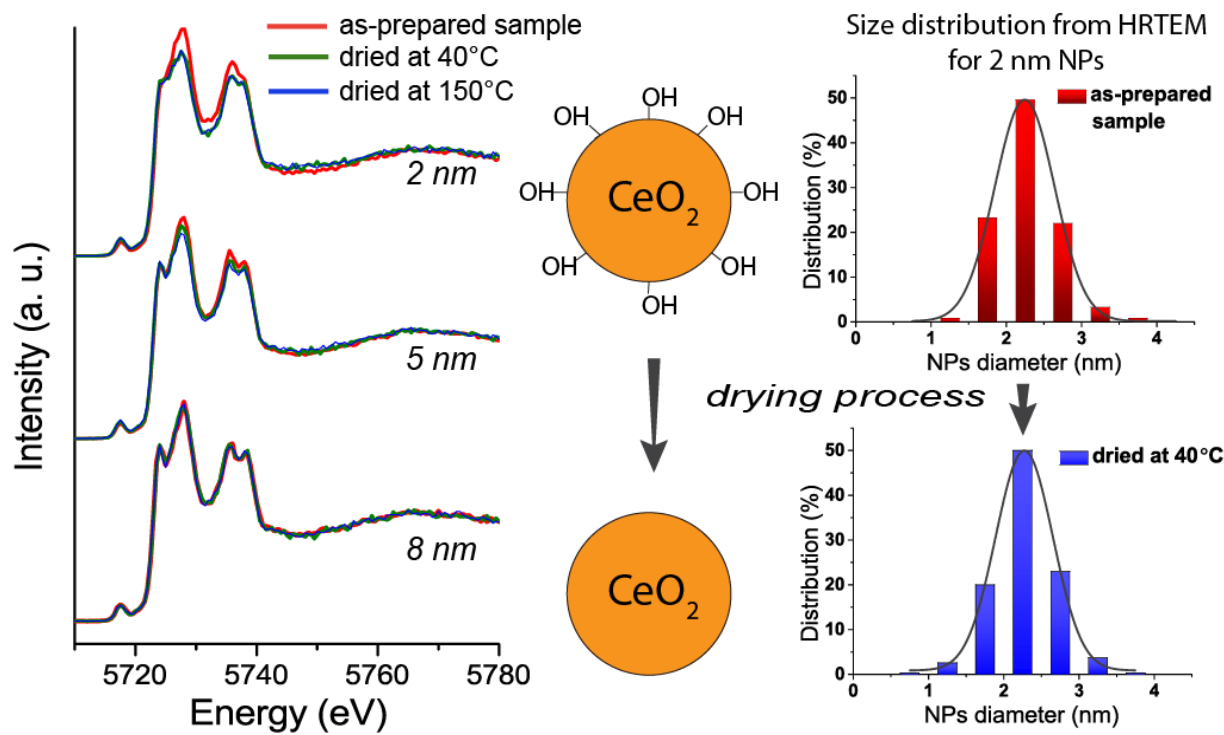

Figure 4. Representative Ce $L_3$ edge HERFD-XAS spectra and size distribution histograms of NPs extracted from HRTEM data for as-prepared NPs and dried NPs at different temperatures.

Understanding the electronic structure changes at the surface is important for many surface-related processes associated with ceria NPs. As noted previously, ceria NPs with a size of less than 10 nm exhibit antioxidant activity scavenging ROS. Such property is commonly attributed to an increase of Ce(III) concentration at the surface, with a decrease of particle size.[50] However, at present, the discussion concerning the mechanisms of ceria biological activity needs further revision to recognize the absence of Ce(III) in the NPs. For instance, Cafun et al.[25] demonstrated that 3 nm ceria NPs, prepared by a wet chemistry technique, catalyse hydrogen peroxide (type of ROS) decomposition. The mechanism of such a reaction remained uncertain, but it was suggested that NP electronic structure changes drastically compared to the bulk $CeO_2$ (mostly due to the Ce-O orbital mixing), since no Ce(III) species were detected in these ceria NPs. We also did not find any indication of the Ce(III) species, but the present study is the first, to our knowledge, to provide the evidence and explore the electronic structure changes of $CeO_2$ NPs associated with the hydroxyl species at the surface of NPs. Due to different surface area to volume ratios for 8, 5 and 2 nm NPs, the surface effect of hydroxyl species should be more pronounced for smaller ceria NPs. Even a very soft thermal treatment of ceria NPs at 40 °C direct measurements of $CeO_2$ NPs antioxidant activity by chemiluminescent method confirm that 2 nm $CeO_2$ NPs do possess higher radical scavenger ability for compare to 8 nm $CeO_2$ NPs (c.f. Fig. S9).

It should be noted that surface hydroxyl species affect not only the biological activity of ceria but also their behavior in several industrial applications. As demonstrated in numerous papers, the presence of water molecules in a catalytic system affects the adsorption properties of ceria oxide.[51] Earlier, it was proposed that $O_2$ cannot adsorb on a perfect $CeO_2$ surface, but strongly binds at oxygen vacancies. With the development of low-temperature catalysts, the proposed mechanism of oxygen sorption has become questionable. There have been extensive experimental studies,[52, 53] which showed that water plays an important role in CO oxidation. According to density functional theory calculations, OH groups can facilitate $O_2$ adsorption on $CeO_2$, even at low OH coverage.[53] Our investigation on the electronic structure of as-prepared and dried ceria NPs clearly demonstrates that OH groups on $CeO_2$ surfaces possess a long-range effect on the $O_2$ adsorption.

## Conclusions

Electronic interactions between the bulk and the surface of NPs control the functionality of nanomaterials. In this study, we have used two large scale synchrotron facilities to study the surface effects of nanoceria at the atomic level. For lanthanide ions, soft X-ray $M_5$ edge spectroscopy allows access to f-orbitals and hard X-ray $L_3$ edge to f- and d-orbitals, which are involved in chemical bonding. Using those techniques, we show that the electronic structure of nanoceria is directly affected by the presence of the hydroxyl groups at the surface. We confirm the entire absence of the Ce(III) oxidation state in $CeO_2$ NPs of different size. Moreover, we study the effect of surface modification by drying NPs of different size at several temperatures. After the temperature treatment the Ce charge remains as Ce(IV); however, the observed difference in the experimentally obtained spectra is attributed to the changes in hydroxylation of the NP surface ($CeO_2$ NPs loose OH groups from their surface). The observed temperature related effect is more pronounced for small NPs, since the surface contributions versus bulk increases for smaller NPs. In this work, we present two distinct theoretical approaches to analyze the experimental evidence, which together confirm our hypothesis of strongly coordinated hydroxyl ligands at the surface of NPs synthesized from aqueous solutions of cerium salts. This finding suggests that changing the cerium ligand bonding of the surface cerium atoms can change its chemical properties and biological activity substantially.

## Conflicts of interest

There are no conflicts to declare.

## Acknowledgements

Authors thank HZDR for the beamtime allocation of project 20-01-800 and P. Glatzel at ID26 beamline of ESRF for providing Ge crystal for the Ce $L_3$ HERFD-XAS measurements. We are also grateful to the "Nanochemistry and Nanomaterials" user facility of the Department of Chemistry of MSU for providing the HRTEM measurements. SAXS measurements were performed using equipment of CKP FMI IPCE RAS.
K.O.K, E.G and S.B. acknowledge support from European Commission Council under ERC grant No. 759696. A.Yu.R, T.V.P. and A.D.K. acknowledge support from RFBR according to the research project No. 18-33-20129. V.K.I. acknowledge support from Russian Science Foundation, grant No. 19-13-00416. S.M.B. acknowledges support from the Swedish Research Council (research grant 2017-06465) and support of travel costs to perform the experiment at SLS from the European Union's Horizon 2020 research and innovation programme under grant agreement n.° 730872, project CALIPSOplus.

## Author Contributions

T.V.P, A.Yu.R. and K.A.D carried out the NP synthesis work. A.V.E., A.E.B. and A.A.S. performed characterizations of the samples using an in-house laboratory equipment. P.V.D. performed synchrotron XRD. K.O.K., A.Yu.R., T.V.P. carried out HERFD-XAS measurements at the Ce $L_3$ edge. E.G, S.B, T.H and S.M.B performed soft X-ray measurements at the Ce $M_5$ edge. S.M.B. and K.O.K. did theoretical simulations of experimental spectra. V.K.I., K.O.K., A.C.S., and S.N.K. were involved in planning and supervision of the work. K.O.K., T.V.P, A.Yu.R, S.M.B., V.K.I., and S.N.K. co-wrote the paper. M.M.S. performed antioxidant activity measurements. All authors discussed the results and contributed to the final manuscript.